\documentclass{article} % For LaTeX2e
\usepackage{iclr2027_conference,times}
\usepackage{amsmath,amsfonts,bm}

\def\eqref#1{equation~\ref{#1}}
\def\1{\bm{1}}

\DeclareMathAlphabet{\mathsfit}{\encodingdefault}{\sfdefault}{m}{sl}
\SetMathAlphabet{\mathsfit}{bold}{\encodingdefault}{\sfdefault}{bx}{n}

\usepackage{hyperref}
\usepackage{url}
\usepackage{booktabs}
\usepackage{graphicx}
\usepackage{capt-of}
\usepackage{subcaption}
\usepackage{array,tabularx,amssymb}
\usepackage{xcolor,colortbl}
\definecolor{tableblue}{RGB}{67,116,163}
\definecolor{tableteal}{RGB}{57,139,119}
\definecolor{tablepurple}{RGB}{132,105,167}
\definecolor{tablebluefill}{RGB}{231,241,249}
\definecolor{tabletealfill}{RGB}{231,246,238}
\definecolor{tablepurplefill}{RGB}{241,235,248}
\definecolor{methodheader}{HTML}{E8EFF5}
\definecolor{methodgroup}{HTML}{EDF1F4}
\definecolor{methodstripe}{HTML}{F5F6F7}

\title{Do Defenses Against LLM Extraction Work Across Attacks? A Lifecycle Benchmark of Black-Box Model Extraction}

\author{Shuze Liu \\
Florida State University \\
\texttt{sl26u@fsu.edu} \\
\And
Kaixiang Zhao \\
Brigham Young University \\
\texttt{kzhao2@byu.edu} \\
\And
Runyang Xu \\
University of Michigan, Ann Arbor \\
\texttt{rx23@fsu.edu} \\
\And
Jingzhi Chen \\
State University of New York at Buffalo \\
\texttt{jingzhic@buffalo.edu} \\
\And
Nathan Wu \\
Wake Forest University \\
\texttt{wuy223@wfu.edu} \\
\And
Yu Wang \\
University of Georgia \\
\texttt{Yu.Wang6@uga.edu} \\
\And
Yushun Dong \\
Florida State University \\
\texttt{yd24f@fsu.edu}
}

\iclrfinalcopy

\begin{document}

\maketitle
% Clear the conference-publication header for this preprint.
\lhead{}

\begin{abstract}
Large language models (LLMs) deployed through text-only APIs face model extraction risks, as adversaries can collect their responses to train surrogates that reproduce their capabilities.
While prior work has developed diverse attacks and defenses, evaluations remain fragmented across access assumptions, model configurations, query budgets, and security objectives, limiting comparability across methods.
To address this gap, we introduce a unified benchmark covering six extraction attacks, ten defenses, and two adaptive attacks that paraphrase or back-translate protected responses before surrogate training.
The benchmark controls model configurations, query data, budgets, and held-out evaluation conditions within each comparison while preserving attack-specific querying and training procedures.
We measure surrogate capability, fidelity to the victim, output quality using Rep-4, and query-budget sensitivity; defenses use their own security metrics paired with surrogate performance.
For the adaptive attacks, we jointly measure provenance-detector scores and the capability and fidelity of surrogates trained on rewritten responses.
The benchmark thus provides a reproducible basis for comparing extraction methods and their interactions with defenses under text-only access.
Code and artifacts are available at \url{https://github.com/sliu11-byte/MEA-Bench}.
\end{abstract}

\section{Introduction}
Many large language models (LLMs) are deployed as proprietary services through text-only APIs \citep{openai2024gpt4technicalreport,birch2023modelleeching}.
Although API access hides model parameters, it exposes the input--output behavior produced through substantial investments in data, computation, and engineering \citep{grattafiori2024llama3,tramer2016apis}.
An adversary can repeatedly query such a model, collect its responses, and use them as supervision for a smaller surrogate \citep{birch2023modelleeching,liang2025lord}.
This functional form of model extraction can reproduce valuable task capabilities without recovering the victim's parameters or training data, creating a direct risk to the confidentiality and economic value of deployed models \citep{zhao2025llmextractionsurvey,ti2025reverseengineering}.

Research on LLM extraction has consequently produced increasingly diverse attacks and defenses.
However, their empirical results are difficult to reconcile because studies vary simultaneously in access assumptions, victim--surrogate configurations, query data, budgets, and evaluation objectives \citep{oliynyk2025evaluation,oliynyk2025factors,zhao2025llmextractionsurvey}.
An attack that appears stronger in one study may therefore benefit from a different model, dataset, or budget rather than a more effective extraction procedure.
Defense claims are even less directly comparable: anti-distillation measures reductions in surrogate performance, provenance and lineage methods use model-level detector scores, and query detectors classify suspicious traffic \citep{savani2025ads,xu2026adfp,yan2026duffin,liu2026mmd}.
Moreover, measuring paraphrasing or back-translation only on protected responses does not show whether a surrogate trained on the rewritten text retains an inherited provenance signal or useful capability \citep{liang2025waterpark,pan2025watermarks}.
Existing benchmarks cover broader privacy threats, knowledge-base leakage, or individual defense families, but, to the best of our knowledge, none compares diverse attacks under matched conditions, tests defenses against their stated objectives, and evaluates adaptive attacks after surrogate training \citep{liu2022mldoctor,huang2024nlpprivacy,qi2026ragbenchmark,jiang2026distillguard}.
As attacks and defenses proliferate, this fragmentation makes reported improvements difficult to attribute and can lead to conflicting conclusions about both extraction risk and defense effectiveness.
This raises a central question: \emph{How can black-box LLM extraction methods be compared under matched conditions while preserving attack procedures and respecting defense-specific objectives?}

Answering this question requires overcoming three limitations in current evaluation practice.
\textbf{(1) Confounded attack comparisons.}
Extraction methods differ in how they acquire queries, construct supervision, and optimize the surrogate \citep{birch2023modelleeching,liang2025lord,chen2026soda,ye2026gad,li2026qedks}.
Because model, data, and budget choices can materially change the measured effectiveness of an attack, comparisons that vary these conditions cannot isolate the contribution of the extraction method itself \citep{oliynyk2025evaluation,oliynyk2025factors}.
Attack comparisons must therefore match the victim and surrogate models, data, and budgets while retaining each method's query-acquisition and training procedure.
\textbf{(2) Different Security Objectives Across Defenses.}
Reducing capability transfer, embedding provenance, verifying lineage, and detecting suspicious queries represent different security objectives rather than interchangeable notions of defense success.
Their evaluations must also expose changes in surrogate capability and fidelity as well as false attribution, which can otherwise make a defense appear effective for the wrong reason \citep{liu2024falseclaims}.
Combining these different measurements into one score obscures what each defense achieves and how it affects surrogate capability and fidelity.
\textbf{(3) Incomplete adaptive evaluation.}
Paraphrasing or back-translation may weaken an embedded signal while also destroying the supervision needed for extraction, or preserve useful supervision while leaving the signal detectable.
Measuring the rewritten text alone cannot distinguish these outcomes or establish whether a defense remains effective against the resulting model.
Adaptive robustness must therefore be evaluated by training a surrogate on rewritten responses and measuring both its detector score and performance.

To address these limitations, we introduce a lifecycle-oriented benchmark that connects attack, defense, and adaptive-attack evaluation under a shared text-only threat model.
To address the first limitation, we implement six representative attacks under common victim--surrogate configurations, query data, and held-out evaluation while retaining their distinct querying and optimization procedures, and separately vary the query budget across different acquisition and training procedures.
To address the second, we organize ten defenses by their intervention points and security objectives and pair each security metric with surrogate capability and fidelity.
To address the third, we insert DIPPER paraphrasing and back-translation before surrogate training and jointly assess the provenance evidence, capability, and fidelity of the resulting surrogates \citep{krishna2023dipper,liang2025waterpark}.
Together, these comparisons isolate extraction methods from model and data choices, test defenses against their stated goals, and determine whether embedded provenance signals survive adaptive rewriting and surrogate training.

Our contributions are summarized as follows:
\begin{itemize}
    \item \textbf{A Taxonomy of Attacks and Defenses by Lifecycle Stage:} We organize extraction attacks, defenses, and adaptive attacks according to their roles and intervention points, providing a common framework for understanding and comparing heterogeneous methods.

    \item \textbf{A Unified Protocol for Evaluating Attacks and Defenses:} We match models, query data, budgets, and evaluation conditions while preserving method-specific procedures, and report results for each defense's stated goal alongside surrogate capability and fidelity.

    \item \textbf{A Comprehensive Evaluation Across Attacks:} Across six attacks, ten defenses, two adaptive attacks, and multiple query budgets, we show that attack rankings depend on the metric, defenses vary in effectiveness across attacks, and weakening provenance signals can reduce surrogate capability and fidelity.
\end{itemize}

\section{Black-Box Extraction Lifecycle and Methods}\label{sec:preliminaries}

\subsection{Black-Box Model Extraction Lifecycle}

\begin{figure}[t]
    \centering
    \includegraphics[width=\textwidth,trim=8bp 130bp 60bp 15bp,clip]{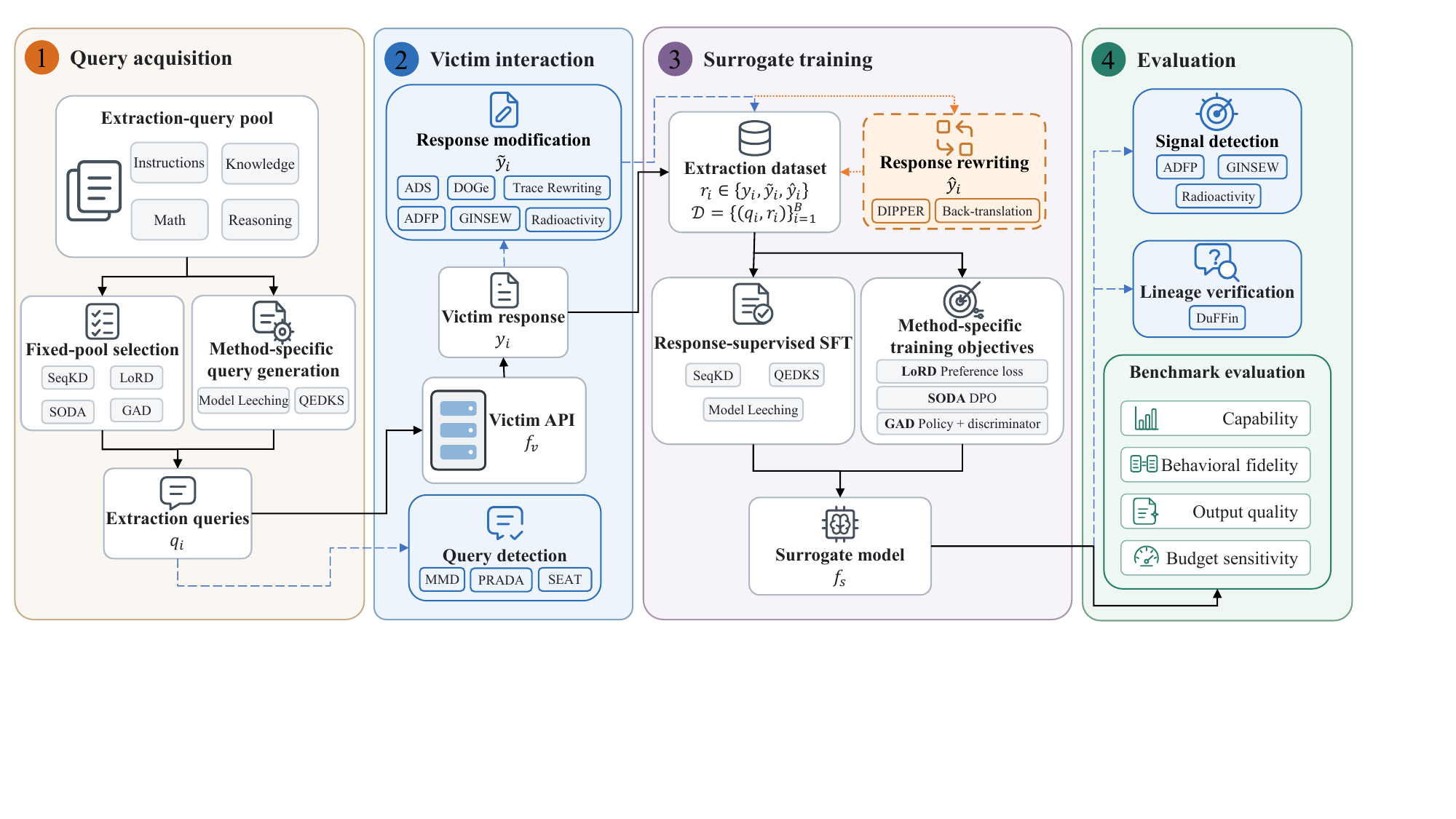}
    \caption{Stages of black-box LLM extraction. Black arrows show extraction flow, blue dashed arrows show defense operations, and orange dotted arrows show adaptive rewriting.}
    \label{fig:extraction_pipeline}
\end{figure}

We consider a victim model $f_{\mathrm{V}}$, also referred to as the teacher, that is accessible through a text-only API, and a surrogate model $f_{\mathrm{S}}$, also referred to as the student, controlled by the attacker.
The defender controls the victim and its released responses, whereas the attacker controls how returned responses are prepared for training and how the surrogate is optimized.
We organize this interaction into the four lifecycle stages shown in Figure~\ref{fig:extraction_pipeline} \citep{oliynyk2025evaluation,zhao2025llmextractionsurvey}.
In \textit{query acquisition}, the attacker selects or constructs the extraction queries $q_i$.
During \textit{victim interaction}, each query is submitted to the victim to obtain a response $y_i=f_{\mathrm{V}}(q_i)$.
In \textit{surrogate training}, the attacker forms an extraction dataset from the queries and the responses used for training and optimizes the surrogate.
Finally, \textit{evaluation} measures the resulting surrogate and any detector outputs used to assess a defense.

The response passed to surrogate training depends on the experimental condition.
In the attack-only setting, the victim response $y_i$ is used directly.
A response-modification defense may instead transform $y_i$ into a protected response $\widetilde{y}_i$.
An adaptive attacker may further paraphrase or back-translate $\widetilde{y}_i$ as $\widehat{y}_i$ before surrogate training \citep{liang2025waterpark}.
Accordingly, for a query budget $B$, the training response $r_i$, extraction dataset $\mathcal{D}$, and resulting surrogate are defined as
\begin{equation}
r_i \in \{y_i,\widetilde{y}_i,\widehat{y}_i\}, \quad
\mathcal{D}=\{(q_i,r_i)\}_{i=1}^{B}, \quad
f_{\mathrm{S}}=\operatorname{Train}(\mathcal{D}).
\end{equation}

Figure~\ref{fig:extraction_pipeline} provides the organizing framework for the attack and defense models in the following subsections.
Extraction attacks may differ in query acquisition and surrogate training, while defenses inspect queries, modify victim responses, or examine trained models.
Adaptive attackers instead rewrite protected responses before surrogate training.
The following subsections describe each method by the lifecycle components it changes.

\subsection{Extraction and Adaptive Attack Models}
\label{sec:attack-models}

The attacker may use public information about the victim's application domain or model family but cannot access its parameters, gradients, hidden states, token probabilities, private training data, or implementation details.
When a defense is present, the attacker does not know its secret keys, watermark rules, or detector thresholds and does not receive the corresponding undefended responses.
The six extraction attacks share the same victim-interaction stage under text-only access: they submit acquired queries to the victim and use the released responses to construct training supervision.
Their defining differences occur during query acquisition and surrogate training, while adaptive attacks additionally transform protected responses after victim interaction and before surrogate training.

\noindent\textbf{Query Acquisition (Stage 1).}
Within the shared text-only access setting, the attacks follow either \textit{fixed-pool selection} or \textit{method-specific query generation}.
SeqKD, LoRD, SODA, and GAD select queries from the shared extraction-query pool \citep{kim2016seqkd,liang2025lord,chen2026soda,ye2026gad}.
Model Leeching applies prompt templates to queries sampled from that pool, whereas QEDKS begins with seed queries and expands them through templates and follow-up questions \citep{birch2023modelleeching,li2026qedks}.
The first group therefore shares the same source and selection process during query acquisition, while the latter methods make query construction part of the attack itself rather than following shared query selection.

\noindent\textbf{Surrogate Training (Stage 3).}
The attacks likewise follow either \textit{response-supervised SFT} or \textit{method-specific training objectives}.
SeqKD and QEDKS train directly on victim responses, while Model Leeching cleans and parses the returned text before applying supervised fine-tuning \citep{kim2016seqkd,birch2023modelleeching,li2026qedks}.
LoRD instead combines victim responses with candidates sampled from the current surrogate for iterative pairwise optimization; SODA constructs teacher--student preference pairs for DPO; and GAD jointly optimizes the surrogate with a response discriminator \citep{liang2025lord,chen2026soda,ye2026gad}.
These methods therefore share the same victim-interaction procedure but differ in how the released responses are converted into supervision and optimized during surrogate training.

\noindent\textbf{Adaptive Attacks (Between Stages 2 and 3).}
We instantiate two adaptive attacks using DIPPER paraphrasing and English--French--English back-translation \citep{krishna2023dipper,liang2025waterpark}.
These transformations target response-embedded provenance evidence while seeking to preserve useful supervision for the downstream surrogate.
Because rewriting occurs after victim interaction, it changes only the supervision passed to surrogate training; the query sequence and victim generation remain fixed.
Both rewriting methods operate on protected responses without clean references, private defense keys, or detector feedback.
The rewritten transcript is then used to train a fresh surrogate under the same SeqKD configuration as the defense-only condition.
We evaluate the resulting model using provenance scores, capability, and behavioral fidelity, since weakening a provenance signal may also remove information that is useful for reproducing the victim's behavior in the extraction setting.

\subsection{Defense Models}
\label{sec:defense-models}

Depending on the defense, the defender can monitor the query sequence, modify generated responses, or compare a trained suspect model with the victim.
The defender possesses any private information required by the defense but cannot control the attacker's processing of returned responses or surrogate training.
We therefore group defenses by what they inspect or modify and by the quantity used to judge whether they succeed.
The observed object is therefore a query, a released response, or a trained model.

\noindent\textbf{Query Detection (Stage 1).}
Before victim inference, MMD, PRADA, and SEAT inspect the queries produced during query acquisition and classify the observed stream as benign or extraction traffic \citep{liu2026mmd,juuti2019prada,zhang2021seat}.
These defenses observe only submitted queries; they cannot use the returned responses, the attacker's training procedure, or the resulting surrogate model.
This restriction distinguishes query detection from response- and model-based defenses at later lifecycle stages.

\noindent\textbf{Response Modification (Stage 2).}
Anti-distillation methods---ADS, DOGe, and Trace Rewriting---seek to reduce the value of the released responses as extraction supervision \citep{savani2025ads,li2025doge,ma2026tracerewriting}.
Provenance methods---ADFP, GINSEW, and Radioactivity---instead embed signals in the released responses that are intended to be inherited by a surrogate trained on them \citep{xu2026adfp,zhao2023ginsew,sander2024radioactivity}.
Both families modify victim responses, but anti-distillation aims to reduce surrogate performance whereas provenance aims to make the trained surrogate detectable during post-training evaluation.

\noindent\textbf{Post-Training Verification (Stage 4).}
Provenance defenses connect two lifecycle points: they embed signals during victim interaction and later apply \textit{signal detection} to the trained surrogate during evaluation.
DuFFin instead performs \textit{lineage verification} only after training, using diagnostic probes to determine whether a suspect model was derived from the victim without modifying the responses released earlier in the lifecycle \citep{yan2026duffin}.
Query detection, anti-distillation, provenance, and lineage verification therefore require different measurements rather than a single defense score.

\section{Experimental Design and Evaluation Procedures}
\label{sec:benchmark-design}

\noindent\textbf{Overview \& Research Questions.}
\label{sec:research-questions}
Following Figure~\ref{fig:extraction_pipeline}, we compare attacks while preserving their distinct query-acquisition and training procedures, evaluate defenses on submitted queries, released responses, or trained models, and apply adaptive rewriting before surrogate training.
Each comparison varies the attack, defense, rewriting, or query budget under study while holding the surrounding experimental conditions fixed.
We report capability and fidelity together with the relevant defense result, such as query classification or a provenance or lineage detector score.
We study four research questions: \textbf{RQ1:} How do different model extraction attacks perform under controlled experimental conditions? \textbf{RQ2:} How well do defenses achieve their objective-specific security goals, and how do they affect surrogate capability and fidelity? \textbf{RQ3:} Can adaptive attacks weaken provenance evidence while preserving surrogate capability and fidelity? \textbf{RQ4:} How does query budget affect attacks with different acquisition and training procedures under otherwise matched conditions?

\subsection{Attack Evaluation Across Lifecycle Stages}
\label{sec:attack-evaluation}

We compare attack effectiveness, protected-response training with and without adaptive rewriting, and query-budget sensitivity while retaining each method's query and training procedure.
Each comparison trains fresh surrogates under the model configuration specified below.

\noindent\textbf{Attack Comparison (Stages 1 and 3; Evaluation at Stage 4).}
All six attacks use a common query budget and shared extraction-query pool while retaining their defining acquisition and training procedures.
Fixed-pool attacks use matched pool prefixes, whereas Model Leeching and QEDKS construct their method-specific queries within the same budget.
Each configuration then trains a fresh surrogate using its method-specific training examples, initialization, and optimization.
The resulting surrogates are evaluated under the same held-out conditions, with capability, behavioral fidelity, and output quality reported separately.

\noindent\textbf{Adaptive-Attack Evaluation (Between Stages 2 and 3; Evaluation at Stage 4).}
Both conditions use the same queries and protected responses.
The \textit{defense-only} condition trains directly on $\widetilde{y}_i$, whereas the \textit{adaptive} condition paraphrases or back-translates each response as $\widehat{y}_i$ before training; both use the same SeqKD configuration.
After training, we compare the surrogate trained on rewritten responses with the one trained directly on protected responses in detector score, capability, and behavioral fidelity under matched evaluation conditions.

\noindent\textbf{Budget Sensitivity (Stage 1; Evaluation at Stage 4).}
We vary the query budget for SeqKD, SODA, Model Leeching, and QEDKS without changing each method's acquisition or training procedure.
SeqKD and SODA use the same fixed-pool transcript at each budget, whereas Model Leeching and QEDKS retain their method-specific acquisition procedures.
Each configuration trains a fresh surrogate and is evaluated under the same held-out conditions.
We then track changes in capability, behavioral fidelity, and output quality as the number of queries increases.

\subsection{Defense Evaluation Across Lifecycle Stages}
\label{sec:defense-evaluation}

Each defense family uses the measurement corresponding to its goal: query classification, surrogate degradation, provenance detection, or lineage separation.
No overall defense score combines these quantities into a single measure.

\noindent\textbf{Query Detection (Stage 1).}
Before victim interaction, each detector is calibrated on benign query batches and then applied to held-out benign and attack batches with matched batch sizes.
The resulting decisions measure how well the detector separates extraction traffic from ordinary use.

\noindent\textbf{Anti-Distillation (Stage 2; Evaluation at Stage 4).}
For each attack--defense pair, the same extraction queries produce clean responses, which ADS, DOGe, or Trace Rewriting then transforms into protected responses.
Fresh surrogates are trained from the clean and protected responses using the same extraction-attack configuration.
After training, anti-distillation is measured by the change in surrogate capability and behavioral fidelity from the matched clean condition to the protected condition.
Because the queries and training procedure are fixed, any performance difference is attributable to response modification.

\noindent\textbf{Provenance (Stage 2; Detection at Stage 4).}
For each provenance defense, we apply its embedding procedure to the clean responses while keeping the queries unchanged.
Surrogates trained on clean and protected responses use the same extraction-attack configuration.
After training, each defense applies its own detector to determine whether the embedded signal survives, while capability and behavioral fidelity measure the resulting changes in surrogate performance.
Across attacks, detector scores are reported separately from capability and fidelity because successful attribution does not imply degraded extraction, or vice versa.

\noindent\textbf{Lineage Verification (Stage 4).}
DuFFin applies the same diagnostic probes to trained extracted surrogates and unrelated control models and measures how well their agreement with the victim separates the two groups.
DuFFin therefore requires trained models rather than matched clean and protected response sets from earlier lifecycle stages.

\subsection{Shared Controls and Experimental Settings}
\label{sec:shared-controls}

\noindent\textbf{Matched Experimental Conditions.}
The variables held fixed depend on the comparison rather than on one protocol shared by all methods.
Queries are matched whenever the method permits.
For a given comparison, we hold the victim and generation settings, query identities where applicable, surrogate configuration, and held-out evaluation conditions fixed unless that component is under study.
Method-specific query acquisition and training procedures are preserved.
Further generation, training, and data-processing details are provided in Appendices~\ref{app:baselines} and~\ref{app:data}.

\noindent\textbf{Query Data and Budgets.}
We construct the extraction-query pool from LMSYS-Chat-1M \citep{zheng2023lmsyschat}, NuminaMath-CoT \citep{numinamath2024}, TriviaQA \citep{joshi2017triviaqa}, MMLU \citep{hendrycks2021mmlu}, GSM8K \citep{cobbe2021gsm8k}, WinoGrande \citep{sakaguchi2020winogrande}, HellaSwag \citep{zellers2019hellaswag}, PIQA \citep{bisk2020piqa}, SciQ \citep{welbl2017sciq}, Social-IQA \citep{sap2019socialiqa}, ARC-Challenge \citep{clark2018arc}, and OpenBookQA \citep{mihaylov2018openbookqa}.
These sources combine open-ended user instructions with mathematical, factual, commonsense, and scientific reasoning tasks, allowing us to evaluate extraction across diverse capabilities.
We use only the input from each source for extraction and evaluate the resulting surrogates on held-out prompts disjoint from the extraction data.
Unless otherwise stated, comparisons use a query budget of $B=1000$; the budget-sensitivity analysis uses $B\in\{100,1000,10000\}$.

\noindent\textbf{Model Configurations.}
We examine two open-weight large-victim/small-surrogate configurations that represent extraction from a hosted model into a trainable local model.
Attack and adaptive-attack experiments use Llama-3.3-70B-Instruct as the victim and Llama-3.1-8B-Instruct as the initial surrogate \citep{grattafiori2024llama3}.
Defense experiments use Qwen2.5-72B-Instruct as the victim and Qwen2.5-7B as the initial surrogate \citep{qwen2024qwen25}.

\noindent\textbf{Evaluation Data and Measures.}
Capability is evaluated on held-out task examples, while behavioral fidelity and output quality use a separate set of open-ended prompts; all evaluation data are disjoint from the extraction queries.
We measure capability using macro-average accuracy, behavioral fidelity using baseline-rescaled BERTScore F1 \citep{zhang2020bertscore}, and output quality using Rep-4.
These dimensions are reported separately rather than combined into an overall ranking.
Defense evaluation additionally uses attack--benign discrimination for query detection, clean--protected performance changes for anti-distillation, native detector scores for provenance, and extracted--unrelated model separation for lineage verification.
Adaptive evaluation reports provenance-detector scores together with the capability and fidelity of the surrogate trained on rewritten responses.
Formal definitions and calibration procedures are provided in Appendix~\ref{app:metrics}.

\section{Experiments}
\label{sec:experiments}

\begin{table*}[t]
\centering
\caption{Attack evaluation at query budget $B=1000$.
Bold marks the best value in each column; Avg. is the six-task macro average, and BERT F1 is baseline-rescaled BERTScore F1.
ARC-C, HellaS., TQA, and WinoG. denote ARC-Challenge, HellaSwag, TruthfulQA MC1, and WinoGrande.}
\label{tab:attack_results}
\scriptsize
\setlength{\tabcolsep}{3pt}
\renewcommand{\arraystretch}{1.08}
\begin{tabularx}{\textwidth}{>{\raggedright\arraybackslash}p{0.16\textwidth}*{9}{>{\centering\arraybackslash}X}}
\toprule
\textbf{Method} & \multicolumn{7}{c}{\textcolor{tableblue}{\textbf{Capability: ACC (\%) $\uparrow$}}} & \textcolor{tableteal}{\textbf{Fidelity}} & \textcolor{tablepurple}{\textbf{Repetition}} \\
\cmidrule(lr){2-8}\cmidrule(lr){9-9}\cmidrule(lr){10-10}
\rowcolor{tablebluefill}
 & ARC-C & HellaS. & MMLU & TQA & WinoG. & GSM8K & \textbf{Avg.} & \mbox{BERT F1 $\uparrow$} & Rep-4 $\downarrow$ \\
\midrule
SeqKD & 51.88 & \textbf{72.20} & 58.64 & 31.82 & 67.25 & 84.84 & 61.10 & 0.337 & 0.124 \\
\rowcolor{methodstripe}
Model Leeching & \textbf{55.97} & 72.18 & \textbf{63.35} & 31.95 & 65.35 & 77.48 & 61.05 & 0.073 & \textbf{0.055} \\
LoRD & 49.23 & 66.37 & 46.40 & 33.41 & 64.56 & 84.46 & 57.41 & 0.289 & 0.188 \\
\rowcolor{methodstripe}
SODA & 51.96 & 72.09 & 58.47 & 32.19 & 66.93 & 85.06 & \textbf{61.12} & 0.338 & 0.122 \\
GAD & 48.38 & 70.80 & 35.10 & 26.19 & \textbf{68.27} & \textbf{86.05} & 55.80 & 0.148 & 0.563 \\
\rowcolor{methodstripe}
QEDKS & 50.26 & 66.82 & 58.52 & \textbf{33.90} & 61.01 & 84.08 & 59.10 & \textbf{0.349} & 0.119 \\
\bottomrule
\end{tabularx}
\end{table*}

\subsection{Extraction Attack Effectiveness}
To answer RQ1, we compare the six attacks at $B=1000$ under the matched conditions in Section~\ref{sec:attack-evaluation} and report accuracy on the evaluation tasks, behavioral fidelity, and output repetition.
From Table~\ref{tab:attack_results}, we make the following observations: (1) From the perspective of capability, SeqKD, Model Leeching, and SODA achieve similar average accuracy, but their advantages vary across tasks.
For example, Model Leeching performs best on ARC-Challenge and MMLU but trails the other two methods on GSM8K, while GAD performs best on WinoGrande and GSM8K despite its lower average accuracy.
These task-dependent rankings show that aggregate accuracy can conceal which capabilities an attack transfers.
Moreover, the more elaborate student-response objectives used by LoRD and GAD do not consistently improve capability over direct teacher-response supervision under the fixed budget, indicating that additional optimization machinery alone does not guarantee broader extraction.
(2) From the perspective of fidelity, the ranking differs from that of capability.
QEDKS attains the highest BERTScore F1 without attaining the highest accuracy, whereas Model Leeching retains near-best capability but has the lowest fidelity.
This contrast is consistent with how they construct training examples: QEDKS trains on complete teacher responses collected through template and follow-up queries, while Model Leeching retains parsed answer text and can therefore preserve task solutions without reproducing the victim's full response form.
Thus, matching the victim's outputs and solving the same tasks represent distinct extraction objectives.
(3) Output repetition provides a third, independent view.
Model Leeching has the lowest Rep-4 but also the lowest fidelity, showing that low repetition measures output diversity rather than faithful imitation; conversely, GAD's high Rep-4 reveals degraded generation quality despite strong results on selected tasks.
In conclusion, no single attack ranks best in capability, fidelity, and output repetition, so all three outcomes are needed for comparison across extraction methods.

\begin{figure*}[t]
\centering
\begin{subfigure}[b]{0.495\textwidth}
\centering
\includegraphics[width=\linewidth]{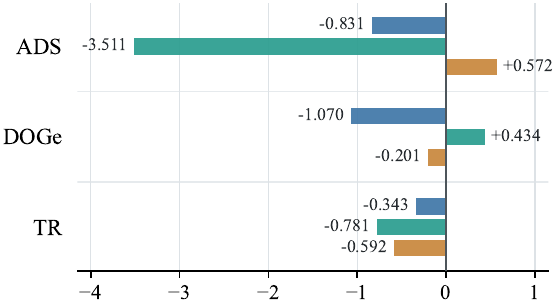}
\caption{Capability: $\Delta$ACC (pp)}
\label{fig:anti_distillation_capability}
\end{subfigure}\hspace{0.005\textwidth}%
\begin{subfigure}[b]{0.495\textwidth}
\centering
\includegraphics[width=\linewidth]{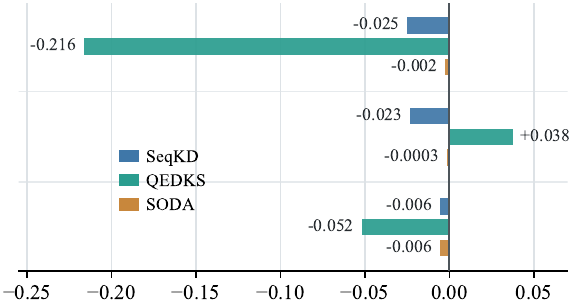}
\caption{Fidelity: $\Delta$BERTScore F1}
\label{fig:anti_distillation_fidelity}
\end{subfigure}
\captionsetup{skip=3pt}
\caption{Anti-distillation effects at $B=1000$.
We report defended-minus-clean ACC (percentage points) and BERTScore F1; negative values indicate suppression, whereas positive values indicate improvement.
Colors distinguish extraction attacks, and TR denotes Trace Rewriting.
Near-zero bars use a minimum visible width; labels give exact values.}
\label{fig:anti_distillation}
\end{figure*}

\begin{figure*}[t]
\centering
\begin{subfigure}[b]{0.326\textwidth}
\centering
\includegraphics[width=\linewidth]{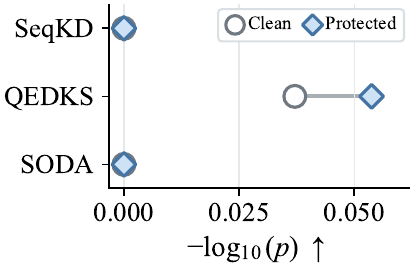}
\caption{ADFP}
\label{fig:provenance_adfp}
\end{subfigure}\hspace{0.006\textwidth}%
\begin{subfigure}[b]{0.326\textwidth}
\centering
\includegraphics[width=\linewidth]{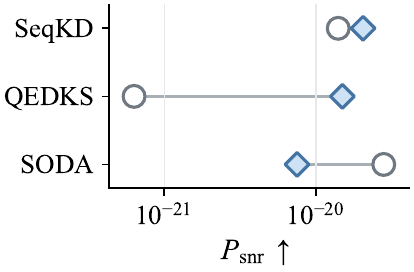}
\caption{GINSEW}
\label{fig:provenance_ginsew}
\end{subfigure}\hspace{0.006\textwidth}%
\begin{subfigure}[b]{0.326\textwidth}
\centering
\includegraphics[width=\linewidth]{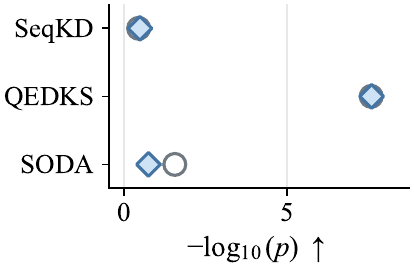}
\caption{Radioactivity}
\label{fig:provenance_radioactivity}
\end{subfigure}
\captionsetup{skip=3pt}
\caption{Provenance-detector scores for surrogates trained on matched clean and protected responses at $B=1000$.
We report ADFP and Radioactivity as $-\log_{10}(p)$ and GINSEW as $P_{\mathrm{snr}}$; larger values indicate a stronger provenance signal.}
\label{fig:provenance_scores}
\end{figure*}

\subsection{Defense Evaluation}
To answer RQ2, we evaluate anti-distillation through capability and fidelity changes, provenance through detector scores, query detection through attack--benign discrimination, and lineage verification through extracted--unrelated model separation.

\textbf{(1) Anti-distillation is attack-dependent.}
As shown in Figure~\ref{fig:anti_distillation} and detailed in Appendix~\ref{app:anti_distillation_results}, ADS suppresses QEDKS but has little effect on SeqKD or SODA, while DOGe is similarly mixed and Trace Rewriting produces more consistent but modest degradation.
These differences are consistent with how the attacks consume protected responses: SeqKD directly imitates them, QEDKS combines teacher supervision with expanded query acquisition, and SODA converts teacher--student outputs into preference pairs.
A modification that disrupts one form of supervision can therefore be attenuated or repurposed by another training objective.

\textbf{(2) Provenance signals do not consistently survive surrogate training.}
Figure~\ref{fig:provenance_scores} and Appendix~\ref{app:provenance_results} show little change in capability or fidelity, but the detector scores vary across extraction attacks.
GINSEW provides its clearest separation under QEDKS, whereas ADFP often returns no signal and Radioactivity can move in the wrong direction.
Thus, embedding provenance signals in released responses does not consistently produce detectable evidence in surrogates subsequently trained on those responses across the evaluated extraction procedures.

\textbf{(3) Query detection depends on the query-generation strategy.}
Appendix~\ref{app:query_detection_results} shows that MMD and SEAT recognize QEDKS traffic, while the fixed SeqKD queries remain largely indistinguishable from benign traffic.
Because these defenses observe queries rather than the attacker's intent or training procedure, QEDKS's generated templates and follow-ups create a more distinctive distribution than queries sampled from a fixed pool.
This contrast suggests that the detectors respond to differences in query distributions caused by the querying procedure, leaving attacks based on natural-looking queries substantially harder to identify.

\textbf{(4) Lineage verification remains unreliable.}
Appendix~\ref{app:lineage_results} shows that DuFFin does not consistently rank extracted students above unrelated controls, indicating that the current probes do not reliably identify models extracted from the victim.
Its success on QEDKS does not extend to SeqKD or SODA, limiting its reliability across extraction procedures.
The score therefore depends on whether extraction transfers victim behaviors covered by the probe set.
Broader probe coverage is therefore needed to compare lineage evidence reliably across attacks.
Such probes should target behaviors preserved across different extraction objectives.

\begin{table*}[t]
\centering
\caption{SeqKD surrogate performance under adaptive attacks at $B=1000$.
ACC is reported in percent; the matched clean baseline has ACC 58.29 and BERTScore F1 0.374.}
\label{tab:adaptive_results}
\footnotesize
\setlength{\tabcolsep}{3pt}
\renewcommand{\arraystretch}{1.10}
\begin{tabularx}{\textwidth}{>{\raggedright\arraybackslash}p{0.20\textwidth}*{6}{>{\centering\arraybackslash}X}}
\toprule
\rowcolor{tablebluefill}
\textbf{Training responses} & \multicolumn{2}{c}{\textcolor{tablepurple}{\textbf{ADFP}}} & \multicolumn{2}{c}{\textcolor{tableteal}{\textbf{GINSEW}}} & \multicolumn{2}{c}{\textcolor{tableblue}{\textbf{Radioactivity}}} \\
\cmidrule(lr){2-3}\cmidrule(lr){4-5}\cmidrule(lr){6-7}
 & ACC & BERT F1 & ACC & BERT F1 & ACC & BERT F1 \\
\midrule
Defense-only & 60.29 & 0.325 & 56.31 & 0.376 & 58.05 & 0.377 \\
\rowcolor{methodstripe}
DIPPER & 56.67 & 0.140 & 56.17 & 0.167 & 56.05 & 0.173 \\
Back-translation & 48.86 & 0.035 & 51.11 & 0.047 & 48.62 & 0.088 \\
\bottomrule
\end{tabularx}
\end{table*}

\begin{figure*}[t]
\centering
\includegraphics[width=0.30\textwidth]{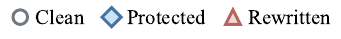}\\[-2pt]
\begin{subfigure}[b]{0.495\textwidth}
\centering
\includegraphics[width=\linewidth]{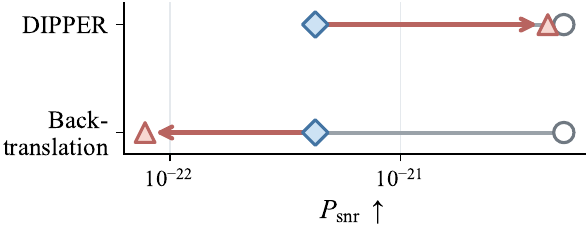}
\caption{GINSEW}
\label{fig:counter_ginsew}
\end{subfigure}\hspace{0.005\textwidth}%
\begin{subfigure}[b]{0.495\textwidth}
\centering
\includegraphics[width=\linewidth]{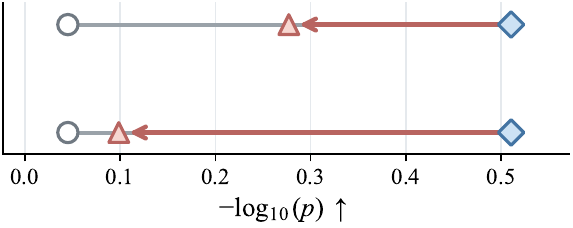}
\caption{Radioactivity}
\label{fig:counter_radioactivity}
\end{subfigure}
\captionsetup{skip=3pt}
\caption{Provenance-detector scores for SeqKD surrogates under response rewriting at $B=1000$.
Rows share clean and protected baselines, and arrows trace scores after rewriting and retraining; larger values indicate stronger evidence.
We omit ADFP because all scores are zero.}
\label{fig:counter_tradeoff}
\end{figure*}
\subsection{Adaptive Robustness}
To answer RQ3, we compare each SeqKD surrogate trained on rewritten responses with its matched surrogate trained directly on protected responses.
Table~\ref{tab:adaptive_results} reports capability and fidelity, Figure~\ref{fig:counter_tradeoff} reports detector scores, and Appendix~\ref{app:adaptive_detailed_results} provides exact values.
From these results, we make the following observations: (1) Rewriting indicates evasion only if protected-response training first raises the detector score above the clean baseline and rewriting then lowers it.
Without clean--protected separation, a lower score cannot demonstrate that an inherited signal was removed.
ADFP produces no inherited signal, while GINSEW lacks positive clean--protected separation; moreover, DIPPER increases the GINSEW score whereas back-translation decreases it, precluding a uniform evasion effect.
Both attacks weaken Radioactivity, with back-translation moving it closer to the clean baseline.
(2) This reduction is not cost-free: both methods reduce fidelity across all three defenses, and back-translation causes the larger capability loss, consistent with greater disruption of embedded provenance signals and responses used for training.
Lowering a detector score benefits the attacker only if the rewritten responses still train a capable and faithful surrogate.
In conclusion, adaptive rewriting produces defense-dependent and partial detector-score reductions while also reducing fidelity, rather than bypassing provenance defenses while preserving extraction quality.

\subsection{Query-Budget Sensitivity}
\begin{figure*}[t]
\centering
\begin{subfigure}[b]{0.326\textwidth}
\centering
\includegraphics[width=\linewidth]{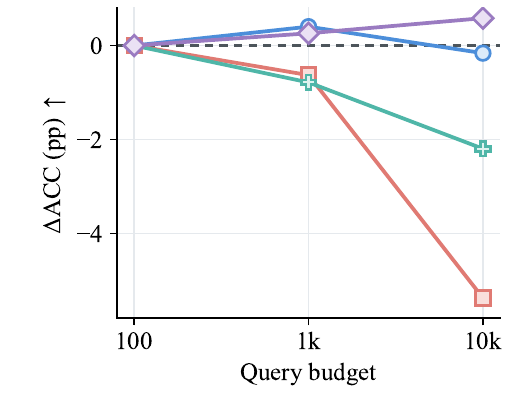}
\caption{Capability}
\label{fig:budget_capability}
\end{subfigure}\hspace{0.006\textwidth}%
\begin{subfigure}[b]{0.326\textwidth}
\centering
\includegraphics[width=\linewidth]{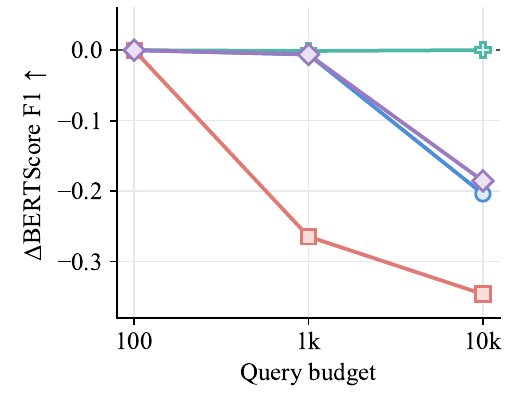}
\caption{Fidelity}
\label{fig:budget_fidelity}
\end{subfigure}\hspace{0.006\textwidth}%
\begin{subfigure}[b]{0.326\textwidth}
\centering
\includegraphics[width=\linewidth]{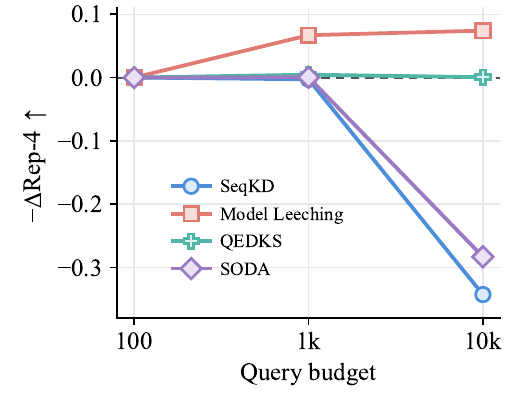}
\caption{Output quality}
\label{fig:budget_quality}
\end{subfigure}
\captionsetup{skip=3pt}
\caption{Query-budget sensitivity of four extraction attacks with distinct query-acquisition and training procedures.
We report changes relative to $B=100$ using $\Delta$ACC and $\Delta$BERTScore F1 and negate $\Delta$Rep-4 so that higher values indicate better outcomes in every panel.}
\label{fig:query_budget}
\end{figure*}

To answer RQ4, we vary the query budget for four attacks while preserving each method's acquisition and training procedure.
Figure~\ref{fig:query_budget} reports changes in capability, fidelity, and output quality relative to $B=100$.

Figure~\ref{fig:query_budget} yields three observations.
(1) Capability does not improve monotonically: SeqKD remains stable and SODA improves slightly, whereas Model Leeching and QEDKS decline, most sharply for Model Leeching from $B=1000$ to $B=10000$.
Thus, additional responses do not ensure transferable supervision.
(2) Fidelity and output quality can decline despite retained capability.
At $B=10000$, both fall sharply for SeqKD and SODA.
Model Leeching loses fidelity as repetition decreases, while QEDKS preserves both despite lower capability; larger budgets can therefore improve one objective while degrading another.
(3) These differences reflect acquisition and training design.
SeqKD and SODA share fixed-pool transcripts, but SODA's preference optimization adds only a modest capability gain and does not prevent large-budget fidelity and output-quality losses.
Parsed-answer supervision separates Model Leeching from the victim's response form, whereas QEDKS preserves response similarity without capability gains.
Overall, budget effects depend on query acquisition and supervision, not query count alone.

\section{Conclusion and Future Work}
\label{sec:conclusion}
We present a lifecycle-oriented benchmark for black-box LLM extraction that compares six attacks, ten defenses spanning four security objectives, and two adaptive attacks under a common text-only threat model.
Each comparison matches the victim--surrogate configuration, query data, and held-out evaluation while preserving method-specific acquisition and training procedures.
No single metric characterizes extraction: attack rankings vary across capability, fidelity, and repetition, and larger budgets do not uniformly improve performance.
Defense effectiveness likewise varies across attacks.
Paraphrasing and back-translation lower some provenance-detector scores but also reduce surrogate capability or fidelity.
These controlled comparisons may change with the victim--surrogate pair, domain, or deployment setting.
Overall, the benchmark shows how querying, training, defenses, and rewriting affect surrogate performance and detection results, and distinguishes changes to responses from effects that persist after training.
Future work will broaden model and language coverage, evaluate stronger adaptive strategies, and develop reliable provenance and lineage-verification protocols.

\subsection*{AI Use Statement}
We used large language models to retrieve and discover relevant literature and to aid in polishing author-written text.
The authors reviewed the retrieved sources, verified the resulting citations against the original publications, and checked all AI-assisted edits for accuracy and consistency with the reported methods and results.
We did not use AI assistance to generate synthetic datasets, prove mathematical claims, or conduct the reported experiments.

\subsection*{Ethics Statement}
Model extraction research has dual-use implications: a systematic evaluation can support stronger defenses, but attack implementations may also facilitate unauthorized imitation of deployed models.
Our study is intended to enable controlled, reproducible analysis of this risk and to clarify the conditions under which existing defenses succeed or fail.
We conduct experiments with public datasets and open-weight models in a controlled research environment; we do not target proprietary production services or collect private user data.
The benchmark does not grant authorization to extract or imitate a model, and its use should comply with model licenses, data licenses, service terms, and applicable law.

\subsection*{Reproducibility Statement}
We specify the procedures for evaluating attacks and defenses and the conditions held fixed in Sections~\ref{sec:attack-evaluation}--\ref{sec:shared-controls}.
Appendices~\ref{app:baselines}--\ref{app:metrics} document the attack and defense implementations, generation and training settings, model configurations, data construction and processing, evaluation metrics, and calibration procedures.
Appendix~\ref{app:complete_results} provides detailed numerical results beyond those reported in the main text.
Code, configuration files, and processed evaluation artifacts are available in the MEA-Bench repository.

\bibliography{iclr2027_conference}
\bibliographystyle{iclr2027_conference}

\appendix

\section{Related Work}
\label{sec:related-work}

\paragraph{LLM Model Extraction.}
Model extraction builds on knowledge distillation, where a student learns to reproduce a teacher's behavior from its predictions \citep{kim2016seqkd}.
Early studies demonstrated imitation of NLP classification and machine translation APIs using query access alone \citep{krishna2020thieves,wallace2020imitation}.
Recent LLM attacks range from template-based response imitation in Model Leeching \citep{birch2023modelleeching}, through student-aware or semi/on-policy optimization in LoRD, SODA, and GAD \citep{liang2025lord,chen2026soda,ye2026gad}, to query-efficient domain extraction in QEDKS \citep{li2026qedks}.
Although surveys have systematized their threat models and techniques \citep{zhao2025llmextractionsurvey,ti2025reverseengineering}, inconsistent query pools, budgets, model configurations, and tasks continue to hinder direct comparison \citep{oliynyk2025evaluation,oliynyk2025factors}.

\paragraph{Defenses and Adaptive Attacks.}
Defenses pursue several distinct goals: ADS, DOGe, and Trace Rewriting make released responses less useful for training a surrogate \citep{savani2025ads,li2025doge,ma2026tracerewriting}; GINSEW, Radioactivity, and ADFP embed provenance signals intended to survive distillation \citep{zhao2023ginsew,sander2024radioactivity,xu2026adfp}; and fingerprinting methods verify model ownership through implanted or behavioral evidence \citep{xu2024instructional,yan2026duffin}.
These approaches build on broader output-watermarking techniques that introduce statistically detectable signals during generation \citep{kirchenbauer2023watermark}.
PRADA, SEAT, and MMD Detector instead seek to recognize extraction from patterns in the attacker's query stream \citep{juuti2019prada,zhang2021seat,liu2026mmd}.
Adaptive attackers can rewrite protected outputs before training: DIPPER studies paraphrasing-based evasion \citep{krishna2023dipper}, while WATERPARK evaluates watermark robustness against paraphrasing, translation, and other transformations \citep{liang2025waterpark}.
Pan et al. further evaluate pre- and post-distillation watermark removal by jointly measuring inherited evidence and student knowledge \citep{pan2025watermarks}.
Recent provenance work also highlights reasoning-specific attribution and the risk that evidence used to claim model ownership is falsely or adversarially reproduced \citep{lv2026reasmark,an2026ditto,liu2024falseclaims}.
Training-time adaptations have also been explored against anti-distillation defenses \citep{allouah2026distillationgame}, but these defense and evasion families are generally evaluated in isolation rather than across an end-to-end extraction lifecycle.

\paragraph{Model Extraction Benchmarks.}
ML-Doctor and an NLP privacy benchmark evaluate model extraction within broader suites of inference or privacy attacks \citep{liu2022mldoctor,huang2024nlpprivacy}, while SecurityNet and subsequent evaluation methodology primarily study public image classifiers and substitute-model stealing \citep{zhang2024securitynet,oliynyk2025evaluation}.
For generative systems, WATERPARK evaluates transformed-text watermark detection \citep{liang2025waterpark}, and \citet{qi2026ragbenchmark} benchmark extraction of protected RAG knowledge.
DistillGuard standardizes the evaluation of output-level LLM distillation defenses under a fixed teacher--student pipeline, but does not compare diverse extraction strategies, provenance and detection objectives, or adaptive attacks \citep{jiang2026distillguard}.
A recent position paper proposes separating active acquisition, fixed-data distillation, detection, and intervention under API constraints, but does not instantiate the proposed benchmark empirically \citep{kurmanji2026distillbench}.
Unlike these complementary but partial settings, our benchmark studies functional extraction from generative LLM APIs and connects diverse attacks, heterogeneous defenses, response transformations used by adaptive attackers, student training, and post-training evaluation in a unified empirical lifecycle.

\section{Implementation and Reproducibility Details}
\label{app:baselines}

\subsection{Shared Generation and Training Settings}

Unless an attack requires online interaction, we generate one victim transcript per query budget and reuse it across attacks so that differences do not arise from stochastic victim outputs.
Victim responses use greedy decoding with temperature $0$, top-$p$ $1.0$, and at most 1536 new tokens.
We use seed 20260701 for query selection and extraction and seed 42 for held-out evaluation.
We preserve each model's chat template, mask prompt tokens from the language-modeling loss, append an end-of-sequence token, and truncate training sequences to 3584 tokens.

We train parameter-efficient surrogates in bfloat16 with gradient checkpointing.
Unless stated otherwise, all trainable surrogate adapters use LoRA \citep{hu2022lora} with rank 16, scaling factor 32, and dropout 0.05.
We do not tune this adapter configuration separately for individual methods or budgets.

\subsection{Attack Implementations}
\label{app:attack_implementations}

We summarize how each attack constructs queries and supervision and optimizes the surrogate in Table~\ref{tab:attack_methods}.

\begin{table*}[t]
\centering
\caption{Extraction attacks compared by how they select queries, construct training examples, and optimize the surrogate. SFT denotes supervised fine-tuning.}
\label{tab:attack_methods}
\small
\setlength{\tabcolsep}{5pt}
\renewcommand{\arraystretch}{1.18}
\begin{tabularx}{\textwidth}{>{\raggedright\arraybackslash}p{0.14\textwidth}>{\raggedright\arraybackslash}p{0.22\textwidth}>{\raggedright\arraybackslash}p{0.29\textwidth}>{\raggedright\arraybackslash}X}
\toprule
\rowcolor{methodheader}
\textbf{Attack} & \textbf{Query selection} & \textbf{Training supervision} & \textbf{Surrogate optimization} \\
\midrule
\rowcolor{methodgroup}
\multicolumn{4}{l}{\textit{Teacher-response SFT}} \\
SeqKD & Fixed pool & Teacher responses & Sequence-level SFT \\
\rowcolor{methodstripe}
Model Leeching & Templated pool queries & Cleaned and parsed teacher responses & SFT \\
QEDKS & Seeds, templates, and follow-ups & Teacher responses & SFT \\
\addlinespace[3pt]
\rowcolor{methodgroup}
\multicolumn{4}{l}{\textit{Student-response-based optimization}} \\
LoRD & Fixed pool & Teacher responses and student candidates & Iterative pairwise optimization \\
\rowcolor{methodstripe}
SODA & Fixed pool & Teacher--student preference pairs & Preference optimization \\
GAD & Fixed pool & Teacher--student discriminator data & Discriminator-guided optimization \\
\bottomrule
\end{tabularx}
\end{table*}

\paragraph{Query selection.}
SeqKD, LoRD, SODA, and GAD draw queries from the same fixed pool, while Model Leeching applies its prompt templates to queries sampled from that pool \citep{kim2016seqkd,birch2023modelleeching,liang2025lord,chen2026soda,ye2026gad}.
QEDKS instead begins with seed queries and uses template-based and follow-up generation to acquire additional domain knowledge within the query budget \citep{li2026qedks}.

\paragraph{Training-data construction.}
SeqKD and QEDKS retain query--teacher-response pairs directly, whereas Model Leeching cleans and parses the teacher output before adding the answer text to its training set \citep{kim2016seqkd,li2026qedks,birch2023modelleeching}.
LoRD augments teacher responses with candidates sampled from the current student, SODA converts teacher and student responses into preference pairs, and GAD uses both sources to construct discriminator supervision \citep{liang2025lord,chen2026soda,ye2026gad}.

\paragraph{Surrogate optimization.}
SeqKD, Model Leeching, and QEDKS optimize the surrogate through supervised fine-tuning on teacher responses \citep{kim2016seqkd,birch2023modelleeching,li2026qedks}.
LoRD performs iterative pairwise optimization using teacher responses and student candidates, SODA applies preference optimization, and GAD alternates updates to the student and a response discriminator \citep{liang2025lord,chen2026soda,ye2026gad}.
This distinction separates attacks that primarily vary query acquisition and response processing from attacks that additionally change the surrogate-training objective.

We report the final optimization settings recorded in the released run manifests in Table~\ref{tab:attack_hyperparameters}.
SeqKD, LoRD, SODA, and GAD use the same fixed victim transcript at each budget.
QEDKS and Model Leeching query the victim online because query construction is part of the attack: QEDKS expands three seed queries into 498 template queries and 499 follow-up queries at $B=1000$, with at most four follow-ups per answer, whereas Model Leeching issues templated prompts and trains on the parsed answer text only.

\begin{table*}[t]
\centering
\caption{Final optimization settings for the extraction attacks. Batch is the per-device batch size and Accum. is the number of gradient-accumulation steps; dashes denote settings not used by LoRD's method-specific optimizer.}
\label{tab:attack_hyperparameters}
\small
\setlength{\tabcolsep}{4pt}
\renewcommand{\arraystretch}{1.12}
\begin{tabularx}{\textwidth}{>{\raggedright\arraybackslash}p{0.14\textwidth}cccc>{\raggedright\arraybackslash}X}
\toprule
\rowcolor{methodheader}
\textbf{Attack} & \textbf{LR} & \textbf{Epochs} & \textbf{Batch} & \textbf{Accum.} & \textbf{Method-specific settings} \\
\midrule
SeqKD & $2\!\times\!10^{-5}$ & 1 & 1 & 32 & Linear schedule; sequence-level SFT. \\
\rowcolor{methodstripe}
Model Leeching & $2\!\times\!10^{-4}$ & 2 & 1 & 16 & Answer-only SFT on parsed online responses. \\
QEDKS & $2\!\times\!10^{-4}$ & 2 & 1 & 16 & Seed/template/follow-up mix; perplexity scheduling disabled. \\
\rowcolor{methodstripe}
LoRD & $3\!\times\!10^{-5}$ & 1 period & -- & -- & Two student candidates; $(\tau_1,\tau_2,\Delta\tau)=(0.8,0.1,0.05)$, $\lambda=0.5$, clipping $0.2$. \\
SODA & $5\!\times\!10^{-6}$ & 1 & 1 & 32 & DPO with $\beta=0.1$; initialized from the matched SeqKD adapter. \\
\rowcolor{methodstripe}
GAD & $1\!\times\!10^{-6}$ & 2 & 1 & 1 & Discriminator LR $10^{-6}$; group size 8; 10 discriminator warm-up steps; KL coefficient $10^{-3}$. \\
\bottomrule
\end{tabularx}
\end{table*}

For LoRD, student candidates are sampled with temperature 1.0, top-$p$ 0.95, and a maximum of 1536 new tokens; we do not initialize LoRD from SeqKD.
SODA forms preference pairs from the victim response and a response sampled from the matched SeqKD surrogate, and then applies one epoch of DPO \citep{rafailov2023dpo} with maximum prompt and sequence lengths of 1024 and 3584.
GAD first performs one SeqKD warm-up epoch at learning rate $5\times10^{-6}$ and then jointly updates generator and discriminator adapters; generator sampling uses temperature 0.8 and top-$p$ 1.0.
We retain the final completed checkpoint for every attack and do not select checkpoints using held-out evaluation results.

\subsection{Defense Implementations and Calibration}

\paragraph{Query detection.}
MMD, PRADA, and SEAT operate before victim inference and classify observed query streams as benign or extraction traffic \citep{liu2026mmd,juuti2019prada,zhang2021seat}.
They receive only the queries available to the model provider and do not inspect the attacker's training procedure or surrogate model.
We encode queries using the \texttt{sentence-transformers/all-MiniLM-L6-v2} sentence-embedding model \citep{reimers2019sentencebert,wang2020minilm}.

\paragraph{Anti-distillation.}
ADS, DOGe, and Trace Rewriting modify the victim's released responses to reduce their usefulness as extraction supervision \citep{savani2025ads,li2025doge,ma2026tracerewriting}.
We train a new surrogate on each protected transcript and compare it with the surrogate trained on matched clean responses.

\paragraph{Provenance and lineage verification.}
ADFP, GINSEW, and Radioactivity embed signals during response generation and subsequently test whether those signals are inherited by the surrogate \citep{xu2026adfp,zhao2023ginsew,sander2024radioactivity}.
DuFFin instead performs post-training lineage verification without modifying the victim responses before extraction \citep{yan2026duffin}.

Defense experiments use the Qwen victim--surrogate pair specified in Appendix~\ref{app:data} and fix the query budget to $B=1000$.
For response-modification and provenance defenses, each matched clean--protected pair retains its attack-specific queries, supervision, initialization, and optimization; only the victim responses differ.
We summarize the final defense settings in Table~\ref{tab:defense_hyperparameters}.

\begin{table*}[t]
\centering
\caption{Final defense and detector settings. Private keys are fixed across embedding and detection, but their numeric values are omitted here.}
\label{tab:defense_hyperparameters}
\small
\setlength{\tabcolsep}{4pt}
\renewcommand{\arraystretch}{1.12}
\begin{tabularx}{\textwidth}{>{\raggedright\arraybackslash}p{0.16\textwidth}>{\raggedright\arraybackslash}p{0.25\textwidth}>{\raggedright\arraybackslash}X}
\toprule
\rowcolor{methodheader}
\textbf{Defense} & \textbf{Configuration} & \textbf{Evaluation or calibration} \\
\midrule
ADS & $\lambda=0.1$, $\epsilon=0.01$, $\tau=0.9$; top-$p$ 0.95 & Proxy gradients are precomputed on held-out prompts and used during protected decoding. \\
\rowcolor{methodstripe}
DOGe & Anti-KD coefficient $3\times10^{-5}$; KD temperature 2.0 & Train the proxy LM head for two epochs with LR $5\times10^{-5}$ and maximum length 2048. \\
Trace Rewriting & Official optimized rewriting prompt; temperature 0.6; top-$p$ 0.95 & Rewrite the complete answer before releasing it to the attacker. \\
\rowcolor{methodstripe}
ADFP & $\gamma=0.5$, context window 2, strength 140 & SHA-256 context seeding with one fixed private key; report $-\log_{10}(p)$. \\
GINSEW & Mark 50\% of eligible tokens; strength 2.0; frequency 16; $\epsilon=0.2$ & Use the same watermark key for generation and detection; report watermark SNR. \\
\rowcolor{methodstripe}
Radioactivity & Maryland, 4-gram context, $\gamma=0.25$, $\delta=2.0$ & Hash-based seeding and the v2 scoring rule; report the detector $p$-value. \\
MMD / PRADA / SEAT & 50 queries per batch; MiniLM-L6-v2 embeddings & Fit on 80\% of benign traffic; lock detector-specific 95\%-tail rules; test on disjoint benign and attack batches. \\
\rowcolor{methodstripe}
DuFFin & MMLU-Pro \citep{wang2024mmlupro} probes from six subject groups & Score the common intersection of 508 valid probes across the victim, base model, and all surrogates. \\
\bottomrule
\end{tabularx}
\end{table*}

For query-stream detection, we construct benign traffic with the same batch size and budget as the attack stream.
The detector receives only queries observed by the victim; follow-up queries produced by QEDKS therefore count toward its budget.
We split benign batches before calibration, fit detector-specific statistics on the calibration portion, and freeze the resulting threshold before evaluating either attacks or held-out benign batches.
MMD estimates its null distribution using 200 samples.
For provenance detection, clean and protected students are evaluated on matched held-out prompts, and the clean score is retained as the reference for each attack--defense pair.

\subsection{Adaptive-Attack Implementations}

We instantiate two representative adaptive attacks against provenance defenses, using DIPPER for paraphrasing and English--French--English back-translation for translation-based rewriting \citep{krishna2023dipper,liang2025waterpark}.
Both methods require only the protected response text and do not use clean responses, private defense keys, detector thresholds, or detector outputs.
We apply adaptive rewriting only to the response text after protected generation and before surrogate training; the queries and $B=1000$ budget remain unchanged.
DIPPER uses \texttt{kalpeshk2011/dipper-paraphraser-xxl} with its T5 tokenizer, lexical diversity 40, order diversity 0, three-sentence rewriting windows, no surrounding document context, top-$p$ 0.75, and maximum length 512.
Back-translation uses \texttt{facebook/seamless-m4t-v2-large} \citep{seamless2023communication} to translate each protected response from English to French and back to English.
Both branches preserve the original query--response pairing and train a fresh SeqKD surrogate with exactly the defense-only training configuration.
For every provenance defense, we separately rerun the detector on the clean, defense-only, and rewritten branches rather than reusing baseline scores across branches.

\subsection{Software, Hardware, and Checkpoint Selection}

We run the experiments under Python 3.10 with PyTorch 2.11.0, Transformers 5.14.1, PEFT 0.15.2, TRL 1.9.2, Accelerate 1.14.0, Datasets 5.0.1, and vLLM 0.26.0.
Cluster jobs load CUDA 12.4.1 and record the resolved accelerator, package freeze, command, configuration, inputs, and output checksums in a run manifest.
We use the final completed adapter checkpoint for evaluation; SODA additionally records and validates its matched SeqKD initialization, and GAD stores the generator and discriminator adapters separately.

\section{Models, Tasks, and Data Processing}
\label{app:data}

\subsection{Victim and Surrogate Models}
\label{app:model_configurations}

We use two fixed victim--surrogate configurations.
Attack and adaptive-attack experiments use \texttt{meta-llama/Llama-3.3-70B-Instruct} as the victim and initialize every surrogate from \texttt{meta-llama/Llama-3.1-8B-Instruct}.
Defense experiments use \texttt{Qwen/Qwen2.5-72B-Instruct} as the victim and \texttt{Qwen/Qwen2.5-7B} as the initial surrogate.
The latter is the base 7B checkpoint rather than its instruction-tuned variant.

Victims are served through OpenAI-compatible text-generation endpoints, and the attacker observes only submitted queries and returned text.
Victim parameters, token probabilities, hidden states, and gradients are unavailable to the attacker.
Surrogates are trained locally, and every comparison starts from the same initial checkpoint within its model family; adapters are trained independently for each attack, defense, budget, or rewriting condition.
We use each checkpoint's native tokenizer and formatting convention and record the resolved model revision in the corresponding run manifest.
Method-specific proxy, discriminator, paraphrasing, and translation models are reported with their implementations in Appendix~\ref{app:baselines}.

\subsection{Extraction Query Pool}
\label{app:query_pool}

We construct a master pool of 100,000 extraction queries by combining general user instructions with task-focused questions.
General instructions account for 80\% of the pool and are drawn from a cleaned version of LMSYS-Chat-1M \citep{zheng2023lmsyschat}.
The remaining 20\% draws on eleven public datasets covering factual knowledge, mathematical reasoning, science, physical and social commonsense, and multiple-choice reasoning.
We report the source split and fixed proportion of each component in Table~\ref{tab:query_pool_composition}.
SeqKD, LoRD, SODA, and GAD directly use prefixes of this pool, while Model Leeching applies its method-specific templates to selected pool queries.
QEDKS instead constructs seed, template, and follow-up queries online; every query received by the victim, including a follow-up, counts toward its budget.

\begin{table*}[t]
\centering
\caption{Composition of the master extraction-query pool.}
\label{tab:query_pool_composition}
\footnotesize
\setlength{\tabcolsep}{3pt}
\renewcommand{\arraystretch}{1.08}
\begin{tabular}{@{}p{0.21\textwidth}p{0.24\textwidth}p{0.27\textwidth}p{0.15\textwidth}p{0.06\textwidth}@{}}
\toprule
\textbf{Dataset} & \textbf{Role} & \textbf{Source split} & \textbf{Processing} & \textbf{Share} \\
\midrule
LMSYS-Chat-1M \citep{zheng2023lmsyschat} & Natural user instructions & default/train & User prompt retained & 80.00\% \\
\rowcolor{methodstripe}
NuminaMath-CoT \citep{numinamath2024} & Mathematical reasoning & default/train & Problem retained & 4.00\% \\
TriviaQA \citep{joshi2017triviaqa} & Factual knowledge & rc.nocontext/train & Question retained & 3.00\% \\
\rowcolor{methodstripe}
MMLU \citep{hendrycks2021mmlu} & Multidomain knowledge & all/auxiliary\_train & Question retained & 2.50\% \\
GSM8K \citep{cobbe2021gsm8k} & Mathematical reasoning & main/train & Problem retained & 2.00\% \\
\rowcolor{methodstripe}
WinoGrande \citep{sakaguchi2020winogrande} & Commonsense coreference & winogrande\_debiased/train & Textual template & 2.00\% \\
HellaSwag \citep{zellers2019hellaswag} & Commonsense completion & default/train & Textual template & 1.50\% \\
\rowcolor{methodstripe}
PIQA \citep{bisk2020piqa} & Physical commonsense & plain\_text/train & Textual template & 1.50\% \\
SciQ \citep{welbl2017sciq} & Science knowledge & default/train & Question retained & 1.00\% \\
\rowcolor{methodstripe}
Social-IQA \citep{sap2019socialiqa} & Social commonsense & default/train & Question retained & 1.00\% \\
ARC-Challenge \citep{clark2018arc} & Science reasoning & ARC-Challenge/train & Question retained & 0.75\% \\
\rowcolor{methodstripe}
OpenBookQA \citep{mihaylov2018openbookqa} & Science reasoning & main/train & Question retained & 0.75\% \\
\bottomrule
\end{tabular}
\end{table*}

\subsection{Prompt Formatting, Parsing, and Deduplication}
\label{app:query_processing}

Only the input side of each source example is retained, and answers or rationales are not provided to the surrogate as gold labels.
The victim instead generates the response used as extraction supervision.
Completion and coreference examples are converted into textual queries, while question-answering and instruction sources retain their input content.
Before query collection, prompts are normalized with Unicode NFKC normalization, case folding, and whitespace collapsing for exact-duplicate removal.
Prompts that exactly match held-out evaluation inputs after the same normalization are removed.
We do not replicate examples to fill the pool.
The resulting query text is formatted using the victim's native interface only when the request is issued; model-specific control tokens are not stored as part of the shared query content.

After preprocessing, the master pool is ordered once using seed 20260701.
The query sets for $B\in\{100,1000,10000\}$ are prefixes of this ordering, so every smaller-budget set is a subset of each larger-budget set.
This construction keeps the source proportions fixed while isolating the effect of increasing the number of victim queries.

\subsection{Held-Out Evaluation Tasks}
\label{app:heldout_tasks}

We evaluate capability on the six public tasks listed in Table~\ref{tab:capability_tasks}.
All evaluations are zero-shot and use the complete stated split.
For multiple-choice tasks, we select the option with the highest conditional log-likelihood; we normalize by continuation length for ARC-Challenge and HellaSwag and use unnormalized scores for MMLU, TruthfulQA MC1, and WinoGrande.
For GSM8K, we allow up to 512 generated tokens, extract the final numeric answer, normalize its numeric representation, and require an exact match.
We compute overall ACC as the unweighted macro average of the six task-level accuracies, so large tasks do not dominate the aggregate.

\begin{table*}[t]
\centering
\caption{Held-out capability tasks used in the final evaluation protocol.}
\label{tab:capability_tasks}
\small
\setlength{\tabcolsep}{5pt}
\renewcommand{\arraystretch}{1.10}
\begin{tabularx}{\textwidth}{>{\raggedright\arraybackslash}p{0.20\textwidth}>{\raggedright\arraybackslash}p{0.20\textwidth}>{\raggedright\arraybackslash}p{0.12\textwidth}>{\centering\arraybackslash}p{0.08\textwidth}>{\raggedright\arraybackslash}X}
\toprule
\rowcolor{methodheader}
\textbf{Task} & \textbf{Configuration} & \textbf{Split} & \textbf{$N$} & \textbf{Task metric} \\
\midrule
ARC-Challenge \citep{clark2018arc} & ARC-Challenge & test & 1,172 & Length-normalized accuracy \\
\rowcolor{methodstripe}
HellaSwag \citep{zellers2019hellaswag} & default & validation & 10,042 & Length-normalized accuracy \\
MMLU \citep{hendrycks2021mmlu} & all & test & 14,042 & Accuracy \\
\rowcolor{methodstripe}
TruthfulQA \citep{lin2022truthfulqa} & multiple choice (MC1) & validation & 817 & MC1 accuracy \\
WinoGrande \citep{sakaguchi2020winogrande} & winogrande\_xl & validation & 1,267 & Accuracy \\
\rowcolor{methodstripe}
GSM8K \citep{cobbe2021gsm8k} & main & test & 1,319 & Strict numeric exact match \\
\bottomrule
\end{tabularx}
\end{table*}

Behavioral fidelity and output quality use a separate set of 3,000 open-ended prompts.
We select three consecutive 1,000-query blocks at positions 10,001--13,000 in the fixed master-pool ordering, immediately after the largest extraction prefix.
These prompts preserve the master pool's source proportions and are disjoint from all extraction sets with $B\leq 10000$.
We issue the same prompts to the victim and every surrogate and retain the complete generated responses for paired fidelity and repetition measurements.
We define the corresponding BERTScore F1 and Rep-4 computations in Appendix~\ref{app:metrics}.

\section{Metrics and Evaluation Protocols}
\label{app:metrics}

\subsection{Capability, Fidelity, and Output Quality}
\label{app:core_metrics}

\paragraph{Capability.}
For each task $t$ in the six-task suite from Appendix~\ref{app:heldout_tasks}, we compute the fraction of correctly answered examples, denoted $a_t$.
We report the unweighted task macro average
\begin{equation}
    \mathrm{ACC}=\frac{1}{6}\sum_{t=1}^{6}a_t.
\end{equation}
We multiply ACC by 100 when reporting percentages.
This aggregation gives every task equal weight regardless of its number of examples.

\paragraph{Behavioral Fidelity.}
We compare each surrogate response with the victim response to the same held-out prompt using BERTScore F1 \citep{zhang2020bertscore}.
We use \texttt{roberta-large}, English baseline rescaling, and the mean over the 3,000 paired responses:
\begin{equation}
    \mathrm{BERTF1}=\frac{1}{N}\sum_{i=1}^{N}
    \operatorname{BERTScoreF1}(y_i^{\mathrm{S}},y_i^{\mathrm{V}}),
    \qquad N=3000.
\end{equation}
Here $y_i^{\mathrm{S}}$ and $y_i^{\mathrm{V}}$ are the surrogate and victim responses.
Baseline rescaling can produce negative values.
We assign a score of one to an empty--empty pair and zero when exactly one response is empty.

\paragraph{Output Repetition.}
We case-fold and tokenize each surrogate response with a Unicode-aware tokenizer that keeps CJK characters as individual tokens.
For a response with at least four tokens, we define
\begin{equation}
    \operatorname{Rep\mbox{-}4}(y)
    =1-\frac{|\operatorname{uniq}(G_4(y))|}{|G_4(y)|},
\end{equation}
where $G_4(y)$ is the multiset of contiguous 4-grams.
We macro-average this value over eligible responses and exclude responses shorter than four tokens.
Lower Rep-4 indicates less within-response repetition; it does not measure correctness or similarity to the victim.

\subsection{Defense-Specific Metrics}
\label{app:defense_metrics}

\paragraph{Anti-Distillation.}
For $m\in\{\mathrm{ACC},\mathrm{BERTF1}\}$, we report
\begin{equation}
    \Delta m=m_{\mathrm{defended}}-m_{\mathrm{clean}}.
\end{equation}
Thus, a negative value means that protected responses reduce surrogate performance relative to the matched clean run.
ACC differences are reported in percentage points.

\paragraph{Provenance.}
We retain each defense's native detector statistic because the scores have different meanings and scales.
ADFP reports $-\log_{10}(p)$, for which larger values indicate stronger fingerprint evidence.
GINSEW reports the Lomb--Scargle signal-to-noise statistic $P_{\mathrm{snr}}$ at the embedded frequency, for which larger values indicate stronger evidence.
Radioactivity reports its detector $p$-value, for which smaller values indicate stronger evidence.
We pair every protected or rewritten student with a clean student trained by the same extraction attack and report the raw scores rather than combining them into a single provenance metric.
For visual consistency, we display Radioactivity as $-\log_{10}(p)$ in Figures~\ref{fig:provenance_scores} and~\ref{fig:counter_tradeoff}, so larger plotted values indicate stronger evidence for every defense.

\paragraph{Query Detection.}
We treat each 50-query group as one decision unit.
For attack batches $\mathcal{A}$ and held-out benign batches $\mathcal{B}$, we compute
\begin{equation}
    \mathrm{TPR}=\frac{\sum_{b\in\mathcal{A}}\mathbb{1}[b\text{ flagged}]}{|\mathcal{A}|},
    \qquad
    \mathrm{FPR}=\frac{\sum_{b\in\mathcal{B}}\mathbb{1}[b\text{ flagged}]}{|\mathcal{B}|}.
\end{equation}

\paragraph{Lineage Verification.}
For DuFFin, we retain probes for which the victim and every evaluated model yield a valid option and compute the victim--model answer agreement
\begin{equation}
    S_{\mathrm{DuFFin}}(f)=\frac{1}{K}\sum_{i=1}^{K}
    \mathbb{1}[c_i(f)=c_i(f_{\mathrm{V}})],
\end{equation}
with $K=508$ common valid probes.
We compute model-level ROC-AUC by comparing the scores of the three extracted students with four negative controls; a tied positive--negative pair contributes $0.5$.

\paragraph{Adaptive Robustness.}
We compare each rewritten condition with its matched defense-only condition using the corresponding provenance statistic, ACC, and BERTScore F1.
We report these quantities separately so that successful signal removal is not conflated with degradation of the rewritten training data.

\subsection{Detection Calibration and Statistical Aggregation}
\label{app:detection_calibration}

For query detection, we use 1,000 benign queries matched to the attack budget.
We reserve 800 queries, or 16 batches, for calibration and hold out the remaining 200 queries, or four batches, for evaluating false positives.
Each attack contributes 1,000 victim-observed queries, forming 20 attack batches.
We calibrate each detector's decision rule on the benign batches and freeze it before evaluating attack and held-out benign batches.
MMD and PRADA use an upper 95th-percentile anomaly cutoff, while SEAT uses the corresponding two-sided benign tails; SEAT additionally calibrates its pair-similarity threshold from benign query pairs.
MMD estimates its null distribution with 200 samples.

All attacks being compared use the same task examples and held-out prompts, and clean--defended and defense-only--rewritten comparisons preserve the query identifiers across conditions.
We use seed 42 for evaluation and detector batching.
We aggregate capability first within each task and then across tasks, while BERTScore F1 and Rep-4 are macro-averaged over held-out responses.
The reported comparisons are therefore paired, descriptive comparisons under a fixed evaluation protocol.

\section{Detailed Experimental Results}
\label{app:complete_results}

The tables in this section provide exact values for experiments summarized in the main text.
When paired with a main-text figure, the table and figure use the same experimental runs: figures emphasize changes or trajectories, whereas tables preserve raw or native-scale measurements for reproducibility.

\subsection{Additional Query-Budget Results}
\label{app:attack_budget_results}

Table~\ref{tab:query_budget_details} reports the exact results for the four attacks in the query-budget analysis at $B=100$ and $B=10000$; their $B=1000$ results are reported in Table~\ref{tab:attack_results}.

\begin{table*}[t]
\centering
\caption{Aggregate attack results at query budgets $B=100$ and $B=10000$.
Rep-4 is the repeated 4-gram rate; bold marks the best value within each budget.}
\label{tab:query_budget_details}
\footnotesize
\setlength{\tabcolsep}{5pt}
\renewcommand{\arraystretch}{1.08}
\begin{tabularx}{\textwidth}{>{\raggedright\arraybackslash}p{0.20\textwidth}*{6}{>{\centering\arraybackslash}X}}
\toprule
\rowcolor{tablebluefill}
\textbf{Method} & \multicolumn{3}{c}{$\boldsymbol{B=100}$} & \multicolumn{3}{c}{$\boldsymbol{B=10000}$} \\
\cmidrule(lr){2-4}\cmidrule(lr){5-7}
 & ACC $\uparrow$ & BERT F1 $\uparrow$ & Rep-4 $\downarrow$ & ACC $\uparrow$ & BERT F1 $\uparrow$ & Rep-4 $\downarrow$ \\
\midrule
SeqKD & 60.70 & 0.343 & \textbf{0.1212} & 60.54 & 0.139 & 0.4641 \\
\rowcolor{methodstripe}
Model Leeching & \textbf{61.68} & 0.337 & 0.1216 & 56.30 & -0.009 & \textbf{0.0475} \\
QEDKS & 59.88 & \textbf{0.350} & 0.1232 & 57.69 & \textbf{0.350} & 0.1227 \\
\rowcolor{methodstripe}
SODA & 60.86 & 0.344 & 0.1223 & \textbf{61.44} & 0.159 & 0.4055 \\
\bottomrule
\end{tabularx}
\end{table*}

\subsection{Anti-Distillation Results}
\label{app:anti_distillation_results}

We report the clean and defended values underlying Figure~\ref{fig:anti_distillation} in Table~\ref{tab:anti_distillation_raw}.
The ACC values use the conditional-log-likelihood evaluation protocol; BERT F1 denotes baseline-rescaled BERTScore F1 against the same undefended teacher.

\begin{table*}[t]
\centering
\caption{Clean and defended surrogate performance under anti-distillation defenses at $B=1000$.
ACC is reported in percent; Figure~\ref{fig:anti_distillation} reports defended-minus-clean changes, for which negative values indicate degradation.}
\label{tab:anti_distillation_raw}
\footnotesize
\setlength{\tabcolsep}{3pt}
\renewcommand{\arraystretch}{1.08}
\begin{tabularx}{\textwidth}{>{\raggedright\arraybackslash}p{0.12\textwidth}*{8}{>{\centering\arraybackslash}X}}
\toprule
\textbf{Attack} & \multicolumn{2}{c}{\textbf{Clean}} & \multicolumn{2}{c}{\textcolor{tableblue}{\textbf{ADS}}} & \multicolumn{2}{c}{\textcolor{tableteal}{\textbf{DOGe}}} & \multicolumn{2}{c}{\textcolor{tablepurple}{\textbf{Trace Rewriting}}} \\
\cmidrule(lr){2-3}\cmidrule(lr){4-5}\cmidrule(lr){6-7}\cmidrule(lr){8-9}
\rowcolor{tablebluefill}
 & ACC & BERT F1 & ACC & BERT F1 & ACC & BERT F1 & ACC & BERT F1 \\
\midrule
SeqKD & 57.62 & -0.019 & 56.79 & -0.044 & 56.55 & -0.042 & 57.28 & -0.024 \\
\rowcolor{methodstripe}
QEDKS & 63.34 & 0.142 & 59.83 & -0.074 & 63.77 & 0.180 & 62.56 & 0.090 \\
SODA & 57.68 & 0.043 & 58.26 & 0.041 & 57.48 & 0.043 & 57.09 & 0.037 \\
\bottomrule
\end{tabularx}
\end{table*}

\subsection{Provenance Results}
\label{app:provenance_results}

We report the capability and fidelity of students trained on clean and protected responses in Table~\ref{tab:provenance_utility}.

\begin{table*}[t]
\centering
\caption{Student capability and fidelity under provenance defenses at $B=1000$.
Parentheses give changes from Clean (ACC in percentage points); positive changes indicate higher surrogate performance, not stronger defense.
BERT F1 is baseline-rescaled BERTScore F1 against the same undefended Qwen teacher.}
\label{tab:provenance_utility}
\footnotesize
\setlength{\tabcolsep}{3pt}
\renewcommand{\arraystretch}{1.08}
\begin{tabularx}{\textwidth}{>{\raggedright\arraybackslash}p{0.16\textwidth}*{6}{>{\centering\arraybackslash}X}}
\toprule
\textbf{Condition} & \multicolumn{2}{c}{\textcolor{tableblue}{\textbf{SeqKD}}} & \multicolumn{2}{c}{\textcolor{tableteal}{\textbf{QEDKS}}} & \multicolumn{2}{c}{\textcolor{tablepurple}{\textbf{SODA}}} \\
\cmidrule(lr){2-3}\cmidrule(lr){4-5}\cmidrule(lr){6-7}
\rowcolor{tablebluefill}
 & ACC $\uparrow$ & BERT F1 $\uparrow$ & ACC $\uparrow$ & BERT F1 $\uparrow$ & ACC $\uparrow$ & BERT F1 $\uparrow$ \\
\midrule
Clean & 57.62 & -0.019 & 63.34 & 0.142 & 57.68 & 0.043 \\
\rowcolor{methodstripe}
ADFP & 57.87 {\scriptsize(+0.25)} & -0.020 {\scriptsize(-0.001)} & 63.89 {\scriptsize(+0.55)} & 0.115 {\scriptsize(-0.028)} & 57.86 {\scriptsize(+0.17)} & 0.041 {\scriptsize(-0.002)} \\
GINSEW & 57.45 {\scriptsize(-0.17)} & -0.020 {\scriptsize(-0.001)} & 63.85 {\scriptsize(+0.51)} & 0.175 {\scriptsize(+0.033)} & 57.69 {\scriptsize(+0.01)} & 0.043 {\scriptsize(+0.000)} \\
\rowcolor{methodstripe}
Radioactivity & 57.55 {\scriptsize(-0.07)} & -0.023 {\scriptsize(-0.004)} & 64.03 {\scriptsize(+0.69)} & 0.177 {\scriptsize(+0.034)} & 57.38 {\scriptsize(-0.30)} & 0.040 {\scriptsize(-0.003)} \\
\bottomrule
\end{tabularx}
\end{table*}

We report the corresponding native detector scores in Table~\ref{tab:provenance_detector_scores}.
These scores preserve each defense's original scale and direction rather than imposing an artificial common metric.

\begin{table*}[t]
\centering
\caption{Native provenance-detector scores for clean and protected students at $B=1000$.
ADFP reports $-\log_{10}(p)$, GINSEW reports $P_{\mathrm{snr}}$, and Radioactivity reports its native $p$-value; arrows indicate stronger evidence.}
\label{tab:provenance_detector_scores}
\footnotesize
\setlength{\tabcolsep}{4pt}
\renewcommand{\arraystretch}{1.08}
\begin{tabularx}{\textwidth}{>{\raggedright\arraybackslash}p{0.17\textwidth}*{6}{>{\centering\arraybackslash}X}}
\toprule
\textbf{Defense} & \multicolumn{2}{c}{\textcolor{tableblue}{\textbf{SeqKD}}} & \multicolumn{2}{c}{\textcolor{tableteal}{\textbf{QEDKS}}} & \multicolumn{2}{c}{\textcolor{tablepurple}{\textbf{SODA}}} \\
\cmidrule(lr){2-3}\cmidrule(lr){4-5}\cmidrule(lr){6-7}
\rowcolor{tablebluefill}
 & Clean & Protected & Clean & Protected & Clean & Protected \\
\midrule
ADFP $\uparrow$ & 0 & 0 & 0.0371 & 0.0538 & 0 & 0 \\
\rowcolor{methodstripe}
GINSEW $\uparrow$ & $1.40{\times}10^{-20}$ & $2.04{\times}10^{-20}$ & $6.33{\times}10^{-22}$ & $1.49{\times}10^{-20}$ & $2.80{\times}10^{-20}$ & $7.51{\times}10^{-21}$ \\
Radioactivity $\downarrow$ & 0.355 & 0.324 & $2.61{\times}10^{-8}$ & $2.54{\times}10^{-8}$ & 0.0275 & 0.177 \\
\bottomrule
\end{tabularx}
\end{table*}

\subsection{Query-Detection Results}
\label{app:query_detection_results}

We report batch-level query-detection results in Table~\ref{tab:query_detection_results}.

\begin{table*}[t]
\centering
\caption{Query detection at \mbox{$B=1000$}.
Bold marks the best values within each query group, including ties; rates use calibrated thresholds.}
\label{tab:query_detection_results}
\footnotesize
\setlength{\tabcolsep}{4pt}
\renewcommand{\arraystretch}{1.08}
% Source: ShuzeLiu11/query_traffic, b1000; counts verified against batch_scores.csv.
\begin{tabularx}{\textwidth}{>{\raggedright\arraybackslash}p{0.20\textwidth}>{\raggedright\arraybackslash}p{0.22\textwidth}*{2}{>{\centering\arraybackslash}X}}
\toprule
\rowcolor{tablebluefill}
\textbf{Queries} & \textbf{Detector} & \multicolumn{2}{c}{\textcolor{tableblue}{\textbf{Detection (\%)}}} \\
\cmidrule(lr){3-4}
 & & TPR $\uparrow$ & FPR $\downarrow$ \\
\midrule
SeqKD & MMD & 0 & \textbf{0} \\
 & PRADA & \textbf{5} & 25 \\
 & SEAT & \textbf{5} & \textbf{0} \\
\addlinespace[1.5pt]
\rowcolor{methodstripe}
QEDKS & MMD & \textbf{100} & \textbf{0} \\
\rowcolor{methodstripe}
 & PRADA & 0 & 25 \\
\rowcolor{methodstripe}
 & SEAT & \textbf{100} & \textbf{0} \\
\bottomrule
\end{tabularx}
\end{table*}

\subsection{Lineage-Verification Results}
\label{app:lineage_results}

We report the DuFFin scores for the seven models evaluated on the 508 shared valid probes in Table~\ref{tab:duffin_scores}.
Across the three extracted students and four negative models, the model-level ROC-AUC is 0.167.

\begin{table*}[t]
\centering
\caption{DuFFin lineage-verification scores; higher values indicate stronger predicted lineage.}
\label{tab:duffin_scores}
\footnotesize
\setlength{\tabcolsep}{4pt}
\renewcommand{\arraystretch}{1.08}
\begin{tabularx}{\textwidth}{>{\raggedright\arraybackslash}X>{\raggedright\arraybackslash}p{0.28\textwidth}>{\centering\arraybackslash}p{0.18\textwidth}}
\toprule
\rowcolor{tablebluefill}
\textbf{Model} & \textbf{Role} & \textbf{Score $\uparrow$} \\
\midrule
Qwen2.5-7B & Negative control & 0.278 \\
\rowcolor{methodstripe}
Qwen2.5-7B-Instruct & Negative control & 0.705 \\
Qwen2-7B-Instruct & Negative control & 0.547 \\
\rowcolor{methodstripe}
Mistral-7B-Instruct-v0.3 & Negative control & 0.360 \\
SeqKD & Extracted & 0.252 \\
\rowcolor{methodstripe}
SODA & Extracted & 0.254 \\
QEDKS & Extracted & 0.467 \\
\bottomrule
\end{tabularx}
\end{table*}

\subsection{Adaptive-Attack Results}
\label{app:adaptive_detailed_results}

We report the detector scores underlying Figure~\ref{fig:counter_tradeoff} in Table~\ref{tab:adaptive_detector_scores}.
The two rewriting branches use the same clean and defense-only detector baselines.
ADFP is zero in every condition and is therefore retained in the table but omitted from Figure~\ref{fig:counter_tradeoff}.

\begin{table*}[t]
\centering
\caption{Provenance signals before and after adaptive response rewriting for SeqKD at $B=1000$.
ADFP reports $-\log_{10}(p)$, GINSEW reports $P_{\mathrm{snr}}$, and Radioactivity reports its native $p$-value; arrows indicate stronger evidence.}
\label{tab:adaptive_detector_scores}
\footnotesize
\setlength{\tabcolsep}{6pt}
\renewcommand{\arraystretch}{1.08}
\begin{tabularx}{\textwidth}{>{\raggedright\arraybackslash}p{0.18\textwidth}>{\raggedright\arraybackslash}p{0.20\textwidth}*{3}{>{\centering\arraybackslash}X}}
\toprule
\rowcolor{tablebluefill}
\textbf{Defense} & \textbf{Rewriting} & \textbf{Clean} & \textbf{Defense-only} & \textbf{Rewritten} \\
\midrule
ADFP $\uparrow$ & DIPPER & 0 & 0 & 0 \\
 & Back-translation & 0 & 0 & 0 \\
\rowcolor{methodstripe}
GINSEW $\uparrow$ & DIPPER & $5.10{\times}10^{-21}$ & $4.27{\times}10^{-22}$ & $4.35{\times}10^{-21}$ \\
\rowcolor{methodstripe}
 & Back-translation & $5.10{\times}10^{-21}$ & $4.27{\times}10^{-22}$ & $7.81{\times}10^{-23}$ \\
Radioactivity $\downarrow$ & DIPPER & 0.901 & 0.309 & 0.528 \\
 & Back-translation & 0.901 & 0.309 & 0.797 \\
\bottomrule
\end{tabularx}
\end{table*}

\end{document}